# Observation of roton-like dispersion relations in acoustic metamaterials

Zhenxiao Zhu[1,#], Zhen Gao[1,# *], Gui-Geng Liu[2#], Yong Ge[3], Yin Wang[3], Xiang Xi[1], Bei Yan[1], Fujia Chen[4], Perry Ping Shum[1], Hongxiang Sun[3,*], Yihao Yang[4,*]

[1]Department of Electrical and Electronic Engineering, Southern University of Science and Technology, Shenzhen 518055, China.

[2]Division of Physics and Applied Physics, School of Physical and Mathematical Sciences, Nanyang Technological University, 21 Nanyang Link, Singapore 637371, Singapore.

[3]School of Physics and Electronic Engineering, Jiangsu University, Zhenjiang, 212013, China.

[4]Interdisciplinary Center for Quantum Information, State Key Laboratory of Modern Optical Instrumentation, ZJU-Hangzhou Global Science and Technology Innovation Center, Zhejiang University, Hangzhou 310027, China.

[#]These authors contributed equally to this work.

[*]E-mail: gaoz@sustech.edu.cn (Z.G.); jsdxshx@ujs.edu.cn (H.S); yangyihao@zju.edu.cn (Y.Y.);

## Abstract

**Roton dispersion relations, displaying a pronounced "roton" minimum at finite momentum, were firstly predicted by Landau and have been extensively explored in correlated quantum systems at low temperatures. Recently, the roton-like dispersion relations were theoretically extended to classical acoustics, which, however, have remained elusive in reality. Here, we report the experimental observation of roton-like dispersions in acoustic metamaterials with beyond-nearest-neighbour interactions at ambient temperatures. The resulting metamaterial supports multiple coexisting modes with different wavevectors and group velocities at the same frequency and broadband backward waves, analogous to the "return flow" termed by Feynman in the context of rotons. Moreover, by increasing the order of long-range interaction, we observe multiple rotons on a single dispersion band, which have never appeared in Landau's prediction or any other condensed matter study. The realization of roton-like dispersions in metamaterials could pave the way to explore novel physics and applications on quantum-inspired phenomena in classical systems.**

The concept of roton, an elementary excitation (quasiparticle) that exhibits dispersion minima at finite momentum, was firstly predicted by Lev Davidovich Landau[1] to explain the unusual thermodynamic properties of superfluid $^4$He, laying the foundation for the development of several areas of modern physics such as Bose-Einstein condensation, superfluidity, phase transition, and quantum field theory. In such condensates, the dispersion relation does not exhibit the usual monotonic growth of energy with momentum in gases, liquids, and solids. Instead, its dispersion relation evolves from linear at low momentum (phonon) to parabolic-like with a minimum (roton) at finite momentum. Following the suggestion by Richard Feynman[2-3], inelastic neutron scattering experiment[4-7] fully confirmed Landau's remarkable intuition. Later, roton-type collective modes have been widely discovered experimentally in different quantum systems at low temperatures, such as quantum spin liquid[8], fractional quantum Hall states[9], two-dimensional Fermi liquid[10], and quantum gas with cavity-mediated long-range interactions[11-12].

Although undoubtedly that the rotons in superfluid are inherently quantum, their unique dispersion relations could in principle also be observed for classical waves such as elastic, acoustic, and electromagnetic waves. One of the candidate platforms to realize the roton-like dispersion relations is acoustic metamaterial[13] that has been widely adopted to study a variety of unusual physic phenomena such as negative refraction[14-16], superlensing[17-18], invisibility cloaking[19-20] and topological protection[21-26]. Indeed, recent theoretical studies[27-28] have suggested that the roton-like dispersion relations can be achieved in classical acoustic metamaterials by introducing beyond-nearest-neighbour interactions. The resulting roton-like dispersion relation has many intriguing features that could manipulate acoustic or electromagnetic waves beyond the standard negative refractive index metamaterials[29], such as broadband backwave propagation, multiple coexisting Bloch modes with different wavevectors at the same frequency, and a large density of states at the roton minimum and maxon maximum with zero group velocity. However, the roton-like dispersion in classical wave systems under ambient conditions has remained elusive to observation to date.

Here, by exploiting long-range interactions, we report the experimental observation of roton-like dispersion relations in nonlocal acoustic metamaterials for the first time.

The fabricated acoustic metamaterial consists of cube resonators and two types of isolated connecting tubes, which play the roles of nearest-neighbour and beyond-nearest-neighbour interactions, respectively. Via direct acoustic measurements, we experimentally observe the roton-like dispersion relations in acoustic metamaterials, confirming Landau's prediction about the roton-like dispersion behaviours in a classical system. Moreover, by increasing the long-range interaction order, more slope inversions can be achieved in the first Brillouin zone, leading to multiple rotons in a single dispersion band, which have never been predicted or observed in any quantum system. We also discuss the impact of structural chirality on roton-like behaviour.

Results

**Observation of the roton-like dispersion relation in a nonlocal acoustic metamaterial**

The three-dimensional (3D) experimental sample of the acoustic metamaterial with the roton-like dispersion relation is fabricated with stereolithography 3D printing technology, as shown in Fig. 1a. It consists of 50 unit cells along $z$ direction. Each unit cell has a cube resonator with side length $d_1$ (see the red region of Fig. 1b). For the convenience of experimental measurement, each cube resonator has a small hole for detecting the acoustic waves in the resonator. These holes are sealed when not in use and have a negligible impact on the dispersion relation of the acoustic metamaterial. The vertical cylindrical tubes with a radius of $r_1$ connecting two nearest-neighbour resonators serve as the nearest-neighbour interactions. The oblique cylindrical tubes with a radius of $r_2$ connecting two third-nearest-neighbour resonators serve as the beyond-nearest-neighbour interactions with an order of $N=3$. The hollow frame with height $d_2$ serves as an auxiliary structure to mediate the third-nearest-neighbour interactions. The oblique cylindrical tubes connect to the auxiliary hollow frame, from which another set of oblique cylindrical tubes are connected to the third-nearest-neighbour of the starting cube resonator to trigger the long-range interaction. The cube resonators, cylindrical oblique tubes, and the auxiliary frames are filled with air and made of hard plastics to ensure that the inner surfaces meet the hard boundary condition.

Besides, the acoustic metamaterial structure is achiral, which exhibits mirror symmetry and inversion symmetry.

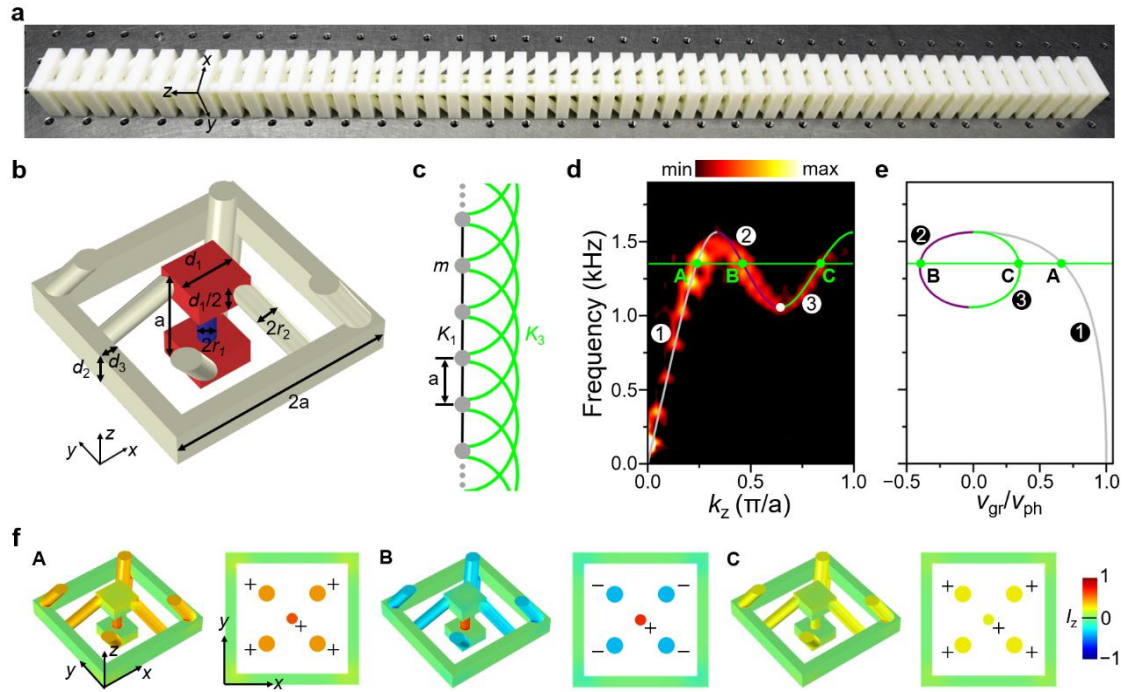

**Fig.1 Experimental observation of the roton-like dispersion relation in an acoustical metamaterial with third-nearest-neighbour interactions. a** Photo of the fabricated experimental sample of the acoustic metamaterial with third-nearest-neighbour interactions. **b** Perspective view of the unit cell along z-direction with periodicity a = 15 mm. Side length of cube resonators (red regions) is $d_1$ = 0.5a. The blue cylindrical connecting tube with radii $r_1$ = 0.08a represents the nearest-neighbour interactions. The beige oblique cylindrical connecting tubes and auxiliary square hollow frame represent the third-nearest-neighbour interactions with geometrical parameters $r_2$ = 0.12a, $d_2$ = 0.32a and $d_3$ = 0.16a. **c** A toy model with third-nearest-neighbour interactions $K_3$ (green lines). The nearest-neighbour interactions $K_1$ are represented by black lines. **d** Measured (red colour) and simulated (colour line) dispersion relations of the architecture in Fig. 1a are shown for propagating acoustic waves along z-direction with wavenumber $k_z$ in the first Brillouin zone. **e** Corresponding group velocity versus frequency, normalized by the phase velocity. The lines with different colours represent the group velocities of different parts of the dispersion relation. **f** Mean energy flux $I_z$ along z-direction corresponding to the three eigenmodes marked as **A**, **B**, and **C** of the dispersion band at the same frequency of 1.35 kHz.

In the experiments, the sound wave generated by a broadband speaker is guided into the waveguide through a small hole at the leftmost cube resonator. A microphone probe is inserted into the other holes at each cube resonator to measure the acoustic pressure signal. This measurement is repeated for all cube resonators of the sample one by one (see Methods). Note that all cylindrical holes are plugged except for the one that opens to measure the acoustic signal. After performing a 1D Fourier transform to the measured acoustic field distributions, we obtain the dispersion relations of the nonlocal acoustic metamaterials (see the colour plotted in Fig. 1d). One can clearly observe the pronounced roton-like dispersion relation--the dispersion band starts from zero frequency and zero momentum with linear growth at low wavevectors, then evolves continuously to reach a maximum, and the dispersion slope (i.e., group velocity $v_g = d\omega/dk$) reverses from positive to negative (see Fig. 1e), displaying a pronounced "roton" minimum (white dot in Fig. 1d) before increasing again. The dispersion relation closely resembles the dispersion curve of the elementary excitations of a Bose superfluid predicted by Landau[1]. The experimental results agree with the simulated counterpart (see the colour curve in Fig. 1d) very well.

The roton-like dispersion relation has several intriguing features, which may find potential applications to manipulate acoustic waves in uncommon ways. First, the "maxon" maximum and "roton" minimum with zero group velocity produce singularities with an extremely high density of states that can be utilized to significantly enhance the wave-matter interactions. Second, the negative slope of roton-like dispersion relation gives rise to broadband backward waves that could lead to acoustic negative refraction and superlensing[14-18] based on a new mechanism of nonlocal effect. Third, at a given frequency, the dispersion curve can support multiple coexisting Bloch modes with different wavevectors, phase velocities, and group velocities.

To understand the origin of the roton-like behaviour in acoustic metamaterials, we plot the simulated mean energy flux along the $z$-direction in a unit cell for three different Bloch modes at the same frequency of 1.35 kHz in Fig. 1f, including two forward modes (**A** and **C**) and one backward mode (**B**) with different wavevectors and group velocities.

For two eigenmodes with positive group velocity (**A** and **C**), the mean energy flux in both the vertical (nearest-neighbour interaction) and oblique connecting tubes (third-nearest-neighbour interaction) is positive. In sharp contrast, for eigenmode with negative group velocity (**B**), the mean energy flux is positive in the vertical connecting tube but is negative in the oblique connecting tubes, which support a backward propagating wave. It is clear that the sum of the two opposite energy fluxes is negative, consistent with the negative group velocity of the dispersion relations shown in Fig. 1d-1e. The two energy fluxes move forward and backward can give rise to a vortex-like behaviour of the energy flux, very similar to Feynman's intuitive appealing picture of rotons which correspond to some sort of stirring motions due to a "return flow"[2].

To further elaborate the experimental results, the practical acoustic metamaterial structure can be translated to a one-dimensional (1D) toy spring-mass model with the acoustic resonator and tubes playing roles of mass and spring (see Fig. 1c), respectively. Here, the black straight lines (green arc lines) represent the nearest-neighbour interactions $K_1$ (the third-nearest-neighbour interactions $K_3$). Newton's equation of motion in this model can be written as

$$m\ddot{u} = K_1(u_{n+1} - 2u_n + u_{n-1}) + K_N(u_{n+N} - 2u_n + u_{n-N}). \tag{1}$$

where $m$ is the mass, $\ddot{u}$ is the acceleration of the mass displacement $u_n$ at lattice site $n$, and $K_N$ is the spring constant representing $N$th-nearest-neighbour interactions, with $N=3$ in our case (see details in Methods). The corresponding dispersion relation is

$$\omega(k) = 2\sqrt{\frac{K_1}{m}\sin^2(\frac{ka}{2}) + \frac{K_N}{m}\sin^2(\frac{Nka}{2})}. \tag{2}$$

By fitting our measured dispersion relation to Eq. (2), we obtain $k_1$=198 $N/m$, $k_3$=454 $N/m$, and $m$=2.59*10$^{-5}$ $kg$ for our experimental sample. The fitted curve is shown in Fig. S1, which is consistent with our measured result.

**Multiple rotons in acoustic metamaterials with higher beyond-nearest-neighbour interaction orders**

Having established the exotic features of the roton-like dispersion in nonlocal acoustic metamaterials with long-range interactions, we discuss the dispersion engineering of the nonlocal acoustic metamaterials by tuning the long-range interaction orders. It has been theoretically predicted[27] that a roton-like minimum occurs if and only if the interactions order is sufficiently high (i.e., $N \geq 3$) and the strength of the beyond-nearest-neighbour interactions is sufficiently large (i.e., $K_N/K_1 > 1/N$). More remarkably, the number of roton-minimum and the negative slope region of the roton-like dispersion can be controlled by tailoring the long-range interaction order. To verify the influence of the order of beyond-nearest-neighbour interactions on the dispersion band, we design and fabricate three acoustic metamaterials with second- ($N=2$), fourth- ($N=4$), and fifth-nearest-neighbour ($N=5$) interactions, respectively. Their unit cells are schematically shown in Figs. 2a, 2e and 2i, respectively, where the vertical cylindrical tubes (blue region) provide the nearest-neighbour interaction ($N=1$) and the oblique cylindrical connecting tubes with the auxiliary hollow frames (beige regions) provide the second- ($N=2$), fourth- ($N=4$) or fifth-nearest-neighbour ($N=5$) interactions independently.

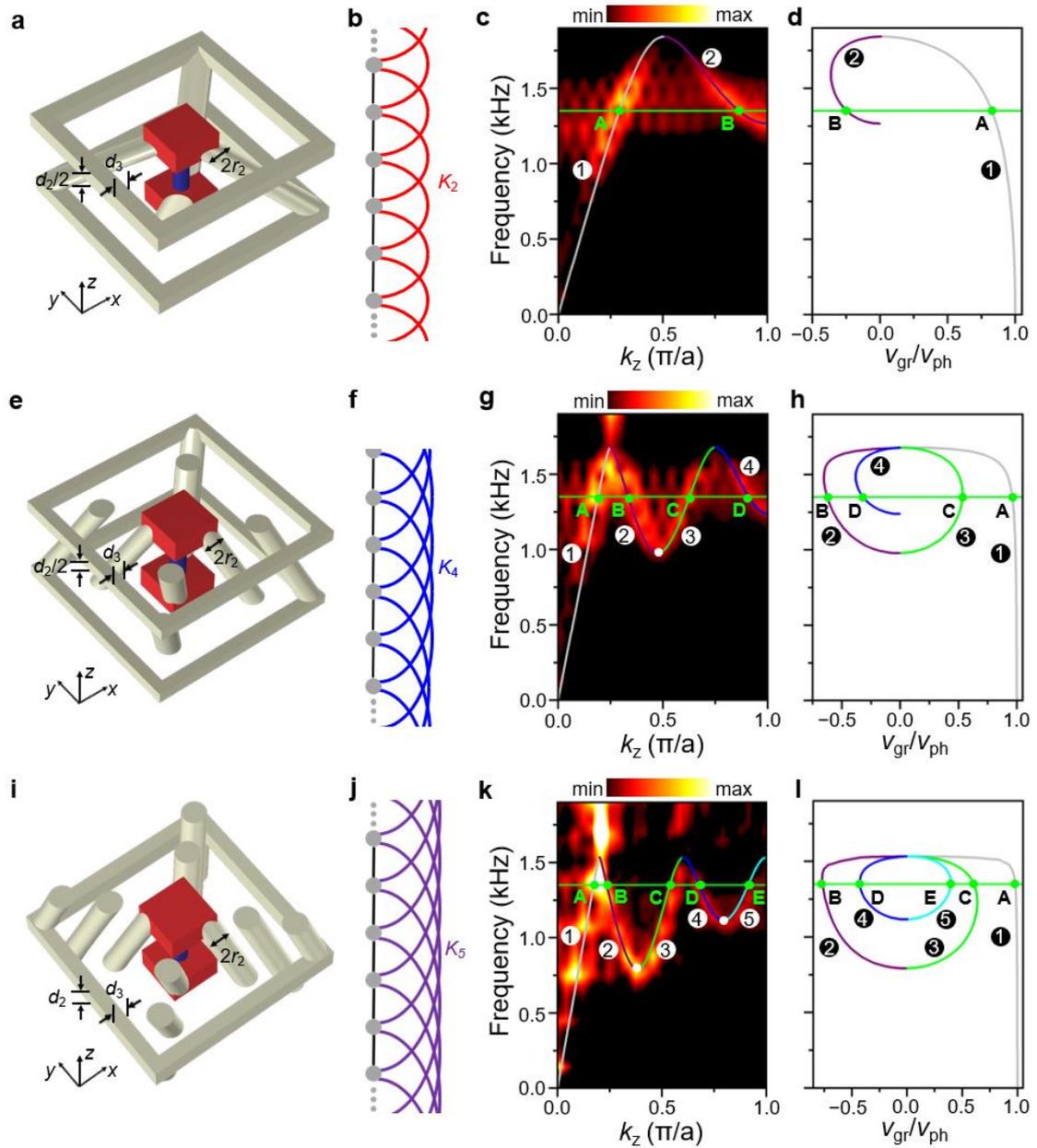

**Fig.2 Dispersion relations in acoustic metamaterials with second, fourth, and fifth-nearest-neighbour interactions, respectively. a** Perspective view of the unit cell that incorporates nearest-neighbour interactions (blue region) and second-nearest-neighbour interactions (beige regions) with geometrical parameters $r_2 = 0.15a$, $d_2 = 0.32a$, and $d_3 = 0.16a$. The other geometrical parameters are the same as those shown in Fig. 1. **b** 1D toy model with the second-nearest-neighbour interactions $K_2$ (red lines). The nearest-neighbour interactions $K_1$ are represented by black lines. **c** Measured (red colour) and simulated (colour line) acoustical dispersion relations and **d** the corresponding group velocity of acoustic metamaterials with the second-nearest-neighbour interactions. **e-h**

Similar to **a-d**, but for the case with the fourth-nearest-neighbour interactions $K_4$ with geometrical parameters $r_2$ = 0.12a, $d_2$ = 0.24a, and $d_3$ = 0.12a. **i-l** Similar to **a-d**, but for the case with the fifth-nearest-neighbour interactions $K_5$ with geometrical parameters $r_2$ = 0.12a, $d_2$ = 0.2a, and $d_3$ = 2a/15. When the order of beyond-nearest-neighbour interactions increases, more slop inversions and local roton minimums (white dot) can be observed in the first Brillouin zone.

For the case of *N=2*, the 1D toy model with second-nearest-neighbour interactions $K_2$ (red arc line) is shown in Fig. 2b. The measured (red colour) and calculated (colour line) band dispersions only exhibit a maxon-maximum, and its group velocity switches sign for only once, as shown in Fig. 2c and 2d, respectively. In this situation, there are only two coexisting eigenmodes with different wavevectors and group velocities at the same frequency. This dispersion relation has no roton minimum and is similar to traditional metamaterial dispersion with a single negative slope region. We then study the case of *N=4*, whose 1D toy model with the fourth-nearest-neighbour interactions $K_4$ (blue arc line) is shown in Fig. 2f. For the case of *N=4*, the measured (red colour) and simulated (colour line) dispersion relations are more complex with the sign of group velocity switches three times and support four different coexisting acoustic eigenmodes at one single frequency, displaying two maxon-maximums and one roton-minimum (white dot), as shown in Fig. 2g and 2h, respectively. Now we continue to increase the long-range interaction order and explore another case with *N=5*, whose 1D toy model with the fifth-nearest-neighbour interactions $K_5$ (purple arc line) is shown in Fig. 2j. The measured (red colour) and simulated (colour line) dispersion relations reverse four times and support five different coexisting acoustic eigenmodes at one single frequency, displaying two maxon-maximums and two roton-minimums (white dots), as shown in Fig. 2k and 2l, respectively. Note that the bright regions above the roton-like dispersion in Figs. 2g and 2k come from a higher band. The mean energy fluxes of different coexisting eigenmodes at the same frequency of 1.35 kHz for *N* = 2, 4 and 5 are shown in Fig. S2. Both the measured and simulated results reveal, as expected, roton behaviour occurs only for beyond-nearest-neighbour interaction order $N \geq 3$ and more slope inversions, and roton minimums can be brought into the first Brillouin zone by increasing the long-range interaction order. Note that multiple classical rotons on a single dispersion curve achieved when $N \geq 5$ have never been experimentally demonstrated in both the quantum and classical-wave systems before.

**Roton-like dispersion relation in a chiral acoustic metamaterial**

Recently, it has been theoretically reported that the roton-like dispersion relations for transverse acoustical elastic waves can be realized in noncentrosymmetric micropolar crystal based on chiral micropolar elasticity theory, where chirality is a necessary condition for roton-like behaviours[30]. Hence, it is natural to ask whether chirality plays an important role in the roton-like behaviours in our airborne acoustic system. Here, we introduce chirality into our acoustic metamaterials with $N$=3 by twisting and doubling the oblique connecting tubes to construct an acoustic metamaterial with four-fold rotational symmetry around the $z$-axis, whose unit cell is schematically shown in Fig. 3a. As both the inversion and mirror symmetries are absent, the newly designed acoustic metamaterial is structurally chiral. Interestingly, both the measured (red colour) and simulated (colour line) dispersion relations of the chiral acoustic metamaterial (Fig. 3b) display a roton-like behaviour which is very similar to that of the achiral acoustic metamaterials in Fig. 1d. The energy flux of the three coexisting modes shown in Fig. 3c further confirmed the similarity between the chiral and achiral acoustic metamaterials with beyond-nearest-neighbour interactions. Therefore, we have experimentally verified that both the chiral and achiral acoustic metamaterials with beyond-nearest-neighbour interactions can support roton-like dispersion relations with similar properties.

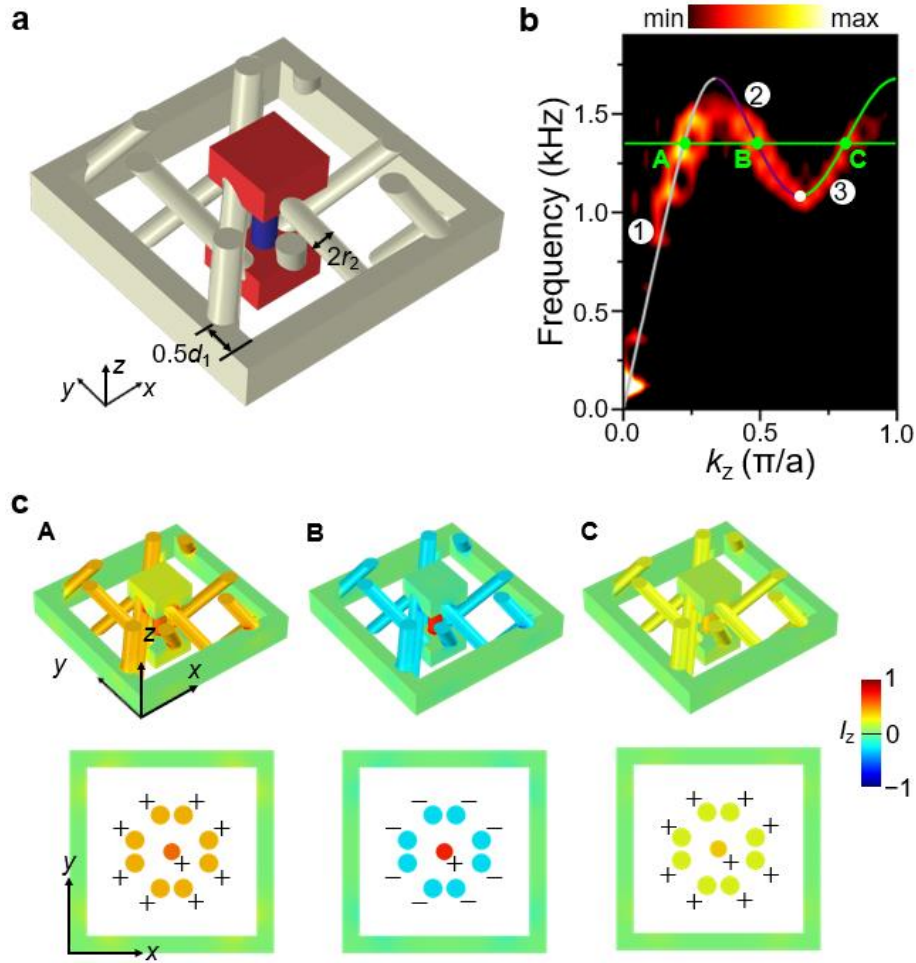

**Fig.3 Roton-like acoustical dispersion relation in a chiral acoustic metamaterial. a** Perspective view of the unit cell of chiral acoustic metamaterials. Similar to Fig. 1 but with "twisted" and doubled oblique tubes to obtain chirality and four-fold rotational symmetry around the $z$-axis. The geometric parameters are the same as those in Fig.1 except for $r_2$ = 0.09a. **b** Measured (red colour) and simulated (colour line) dispersion relations of the chiral acoustic metamaterials. **c** Mean energy flux $I_z$ along the $z$-direction for the three eigenmodes **A**, **B** and **C** at the same frequency of 1.35 kHz.

## Discussion

In conclusion, we have designed and experimentally demonstrated the long-sought roton-like dispersion relations in nonlocal acoustic metamaterials with beyond-nearest-neighbour interactions. With increasing the order of beyond-nearest-neighbour interactions, multiple slope inversions and classical rotons can be achieved on a single dispersion band which have no quantum equivalent. Moreover, we also experimentally verify that both chiral and achiral acoustic metamaterials with long-range interactions can support roton-like dispersions. The realization of roton-like dispersions opens up multiple avenues for explorations of metamaterials, including broadband negative refractive index media and nonlocal metamaterials with arbitrary dispersion relations. Our work points to a direction beyond the nearest-neighbour interactions between the constituent elements of a physical system and offers a powerful platform to engineer the dispersion relations that could enable the observation of unusual physical phenomena and the abundant applications beyond traditional negative-index metamaterials. Although demonstrated in acoustic metamaterials, the same principle can be extended to other classical-wave systems such as photonics, plasmonic, and mechanical systems.

**Methods**

**Numerical simulations.** All the simulations are performed in the pressure acoustic module of COMSOL Multiphysics. In all simulations, the background medium air is modelled with a density of 1.29 kg/m$^3$ and a speed of sound of 340 m/s. The plastic stereolithography can be considered as a hard boundary during the simulations due to the substantial acoustic impedance contrast compared with air. For calculating the dispersion relations, periodic boundary conditions are set in the $z$ direction, and other boundaries are considered as hard acoustic boundaries.

**Experimental details.** The samples are fabricated with an additive manufacturing technique (that is, stereolithography) with a rein of a thickness of 5 mm and 50 periodic unit cells stacked along the $z$-direction. The materials are photosensitive resin with a modulus of 2880 MPa and a density of 1.10gcm$^{-3}$. The structures can be viewed as rigid walls for the acoustic waves that propagate in the acoustic metamaterials due to the large impedance mismatch between air and the material. A small hole ($r$=1.2 mm) is located on the centre of each cube resonator for excitation and detection. When not in use, they are blocked by plugs. In the experiment, a broadband sound signal is launched by a balanced armature speaker guided into the sample through a small hole on the centre cube resonator to excite the acoustic modes. Another microphone (Brüel & Kjær Type 4961, of about 3.2 mm radius) in a sealed sleeve with a tube of 1 mm radius and 250 mm length inserts into each cube resonator through the small hole to collect the acoustic signal one by one. The measured data is processed by a Brüel & Kjær 3160-A-022 module to extract the frequency spectrum with 2 Hz resolution. Spatial Fourier transforms are applied to the complex acoustic field distributions to obtain the measured dispersion relations.

**Acknowledgements**

This work was sponsored by Southern University of Science and Technology for SUSTECH Start-up Grant (Y01236148), Young Thousand Talent Plan of China and the National Natural Science Foundation of China under grant number 12104211. The work at Zhejiang University was sponsored by the National Natural Science Foundation of China under grant number 62175215. H.S. acknowledged the support of the National Natural Science Foundation of China (Grant Nos. 11774137 and 12174159), and the State Key Laboratory of Acoustics, Chinese Academy of Science under grant No. SKLA202016.

**Author contributions**

Y.Y. and Z.G. initiated the project. Z.Z. designed the structures and did the numerical simulations with help from G.L. and X.X. Z.G. Z.Z., H.S., Y.B., and F.C. fabricated the samples. H.S. and Y.G. performed the measurements. Z.Z., Z.G. and Y.Y. analyzed the data and wrote the manuscript. P. S. helped to revise the manuscript. Z.G. and Y.Y.

supervised the project.

**Competing financial interests**

The authors declare no competing financial interests.

**Data availability**

The data that support the findings of this study are available from the corresponding author upon reasonable request.